\documentclass[aps,prl,twocolumn,superscriptaddress,showpacs]{revtex4}

\usepackage{graphicx}
\usepackage{amssymb}
\usepackage{amsmath}
\usepackage{amsfonts}
\usepackage{mathrsfs}

\begin{document}

\title{Quantitative approach to
facilitated diffusion with intersegmental jumping}

\author{Michael A. Lomholt}
\affiliation{MEMPHYS - Center for Biomembrane Physics, Department of
Physics and Chemistry, University of Southern Denmark, Campusvej 55,
5230 Odense M, Denmark}
\author{Bram van den Broek}
\affiliation{Department of Physics and Astronomy, Faculty of Sciences,
VU University, De Boelelaan 1081, 1081 HV Amsterdam, The Netherlands}
\affiliation{Physics of Life Processes, Leiden Institute of Physics,
Leiden University, Niels Bohrweg 2, 2333 CA Leiden, The Netherlands}
\author{Svenja-Marei J. Kalisch}
\affiliation{Department of Physics and Astronomy, Faculty of Sciences,
VU University, De Boelelaan 1081, 1081 HV Amsterdam, The Netherlands}
\author{Gijs J.L. Wuite}
\affiliation{Department of Physics and Astronomy, Faculty of Sciences,
VU University, De Boelelaan 1081, 1081 HV Amsterdam, The Netherlands}
\author{Ralf Metzler}
\affiliation{Physics Department, Technical University of Munich,
D-85747 Garching, Germany}

\begin{abstract}
We study the impact of DNA coiling on the search rate of proteins moving
along the DNA contour interspersed by three-dimensional (3D)
bulk excursions. When the DNA is
coiled proteins performing short 3D hops along a DNA segment can be captured
by foreign DNA segments that have looped back close to the original segment.
These intersegmental jumps are shown to enhance the search rate for a specific
site on the DNA by lowering the tendency to resample previously visited sites
on the DNA. The model developed here offers a quantitative description
of recent single molecule experiments on facilitated diffusion of restriction
enzymes EcoRV.
\end{abstract}

\date{\today}

\pacs{82.39.-k, 02.50.Ey}

\maketitle

A single stochastic search mechanism is usually not efficient at optimizing
a search process. Thus Brownian motion, whose fractal dimension is $d_f=2$,
in 1D and 2D leads to significant oversampling in the
sense that a given site is revisited many times \cite{hughes,lomholt_pnas}.
L{\'e}vy motion with a jump length distribution $\lambda(x)\simeq|x|^{-1-
\alpha}$ ($0<\alpha<2$) has $d_f=\alpha$ and thus a comparatively reduced
degree of oversampling. However, the first arrival of L{\'e}vy motion is
hampered by leapovers, i.e., frequent overshoots of the target \cite{koren}.
In 3D both Brownian
and L{\'e}vy motions are inefficient at locating a small target. While typical
chemical reactants in a liquid environment have no choice but to passively
diffuse until mutual encounter \cite{smoluchowski} intermittent models
combine different search mechanisms \cite{lomholt_pnas,benichou_pccp}.

An important example for intermittent search is facilitated diffusion of
DNA binding proteins searching for their specific target binding site on a
DNA molecule. In the Berg-von Hippel model \cite{bvh} proteins intermittently
diffuse in 3D bulk or along the 1D DNA chain \cite{igor}, the latter mediated
by non-specific binding (NSB) \cite{audun}.

Previous approaches to facilitated protein diffusion consider bulk exchange
for a straight cylindrical chain \cite{berg,REM}; they consider scaling
properties in different chain configurations \cite{hu} without quantitative
comparison of the search rates between these; or they reformulate the problem
with uncorrelated bulk excursions \cite{coppey,lomholt,halford,slutsky}. Here
we relate a multiparticle picture to single protein jumps along the DNA contour
allowing us to consider DNA as a fluctuating chain. Comparison to recent
single DNA measurements \cite{gijs} demonstrates that our model quantitatively
describes the effect of DNA coiling.

\emph{General model.}
In our description of the target search process, we use the density per length
$n(x,t)$ of proteins on the DNA as the relevant dynamical quantity ($x$ is
the distance along the DNA contour). We include 1D diffusion along the DNA
with diffusivity $D_{\mathrm{1d}}$, protein dissociation with rate
$k^{\rm ns}_{\mathrm{off}}$ and (re)adsorption after diffusion through the
bulk fluid with diffusion constant $D_{\rm 3d}$. The dynamics of $n(x,t)$
is thus governed by 
\begin{eqnarray}
&&\hspace*{-0.4cm}\frac{\partial n(x,t)}{\partial t}=\left(D_{\rm 1d}\frac{\partial^2}{
\partial x^2}-k^{\rm ns}_{\rm off}\right)n(x,t)-j(t)\delta(x)+G(x,t)\nonumber\\
&&+k^{\rm ns}_{\rm off}\int_{-\infty}^\infty d x'\int_0^t d t'\;W_{\rm bulk}
(x-x',t-t')n(x',t').
\label{eq:themodel}
\end{eqnarray}
Here, $j(t)$ is the flux into the target site located at $x=0$, $W_{\rm bulk}
(x-x',t-t')$ is the joint probability that a protein returns to the point $x$
at time $t$ after leaving the DNA at $x'$, $t'$ for a bulk excursion. $G(x,t)$
is the flux onto the DNA of proteins that have not previously been bound to
the DNA. $j(t)$ is determined by imposing the boundary condition $n(x=0,t)=0$
at the target. To consider $W_{\rm bulk}$ as homogeneous in space and time
we made two assumptions that we found to be valid for the system studied
in \cite{gijs}. First, that
the DNA can be treated as being infinitely long when considering the search rate \cite{REMinf}. Second, that
the coiled DNA conformation fluctuates quickly such that subsequent excursion
distances $x-x'$ can be treated as independent.

To proceed, we Laplace and Fourier transform Eq.~(\ref{eq:themodel}):
\begin{equation}
n(q,u)/W(q,u)-n_0(q)=G(q,u)-j(u)
\label{eq:Laplace}
\end{equation}
with $n(q,u)=\mathscr{L}\{n(q,t)\}$; $n(q,t)=\mathscr{F}\{n(x,t)\}$;
$n_0(x)=n(x,t=0)$, and
\begin{equation}
W(q,u)=1/\{u+D_{\rm 1d} q^2+k^{\rm ns}_{\rm off}[1-W_{\rm bulk}(q,u)]\}.
\end{equation}
We rewrite these equations in the form
\begin{equation}
n(q,u)=n^{\rm ns}(q,u)-j(u)W(q,u)
\end{equation}
where the density $n^{\rm ns}(q,u)=W(q,u)[G(q,u)+n_0(q)]$ is the solution of
the diffusion problem in absence of the target, i.e., when only NSB between
proteins and the DNA occurs. In position space at $x=0$ one obtains
$j(u)=n^{\rm ns}(x=0,u)/W(x=0,u)$.

In this study we focus on the long time behavior of $j(t)$ for which the
density $n^{\rm ns}(x,t)$ will reach an equilibrium value $n^{\rm ns}_{\rm
eq}$. Since, by Tauberian theorems, large $t$ correspond to small $u$ we
study the $u\to0$ behavior of $j(u)$, with $n^{\rm ns}(x,u)\sim n^{\rm ns}_
{\rm eq}/u$. For $W_0(x=0,u)$ we assume that the limit $W(x=0,u=0)$ is finite
and non-zero; this is true when the distance distribution of bulk excursions
is sufficiently long tailed as fulfilled for the $W_{\rm bulk}$ encountered
below. In practice, being interested in the \emph{first arrival\/} at the
target, we assume that the protein concentration is sufficiently dilute such
that this arrival happens after reaching $n^{\rm ns}_{\rm eq}$ (`rapid
equilibrium') but still with many searching proteins. Thus $j(u)\sim k_{\rm
1d}n^{\rm ns}_{\rm eq}/u$, and therefore at long times the stationary current
$j_{\rm stat}\sim k_{\rm 1d}n^{\rm ns}_{\rm eq}$ into the target is reached.
Here we introduced $k_{\rm 1d}^{-1}=W(x=0,u=0)$:
\begin{equation}
\label{int}
k_{\rm 1d}^{-1}=\int_{-\infty}^\infty \frac{d q}{2\pi}
\frac{1}{D_{\rm 1d}q^2+k_{\rm off}(1-\lambda_{\rm bulk}(q))}.
\end{equation}
$\lambda_{\rm bulk}(x)=W_{\rm bulk}(x,u=0)$ is the distribution of relocation
lengths along the DNA after a single 3D excursion.

We express our results in terms of the association rate
\begin{equation}
k_{\rm on}=j_{\rm stat}/n_{\rm bulk}=k_{\rm 1d}K_{\rm ns}
\end{equation}
to the target where $n_{\rm bulk}$ is the density of unbound proteins in
the bulk (at non-specific equilibrium). The constant of NSB per length of
DNA is $K_{\rm ns}=n^{\rm ns}_{\rm eq}/n_{\rm bulk}$.
In terms of the total
volume concentration of proteins ${n}_{\rm total}={n}_{\rm bulk}+l^{\rm total}
_{\rm DNA}n^{\rm ns}_{\rm eq}$, where $l^{\rm total}_{\rm DNA}$ is the total
length of DNA divided by the volume of the entire system, we have $j_{\rm stat}\sim k_{\rm on}n_{\rm total}/(1+K_{\rm ns}l_{\rm DNA}^{\rm total})$.
The rate $k_{\rm on}$ is related to the mean first arrival time for the steady
state association. Namely $k_{\rm on}{n}_{\rm bulk}$ is the probability per
time for protein association with the target. Thus the (survival) probability
that no protein has arrived at the target yet is $P_{\rm surv}(t)=\exp(-k_{\rm
on}{n}_{\rm bulk} t)$, and the average target search time is $T=1/(k_{\rm on}
{n}_{\rm bulk})$. The flux $j(t)$ being linear in the protein concentration
$n_{\rm total}$ implies that for sufficiently low $n_{\rm total}$ the steady
state is reached well before the first protein actually binds to the specific
target (note that \emph{in vivo\/} protein concentrations can be as low as
nanomolar). 

\begin{figure}
\includegraphics[width=6.8cm]{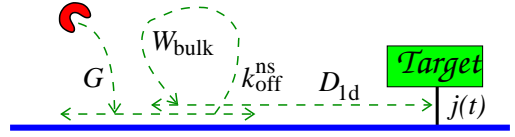}
\caption{Schematic of the search mechanisms in Eq.~(\ref{eq:themodel}) on a
piece of stretched DNA. Note the tendency to revisit sites when immediate rebindings and thus short hops are likely.}
\label{fig:straight}
\end{figure}

\emph{Straight rod configuration (Fig.~\ref{fig:straight}).}
Consider first stretched DNA corresponding to a cylinder of radius $r_{\rm
int}$ (the range of non-specific interaction), with NSB reaction
rate $k^{\rm ns}_{\rm on}$ at the boundary: the flux of proteins, per length of
DNA, onto the DNA is $k^{\rm ns}_{\rm on}$ times the bulk concentration next
to the DNA. This implies that at equilibrium of NSB, $n^{\rm ns}_{\rm eq}k^{
\rm ns}_{\rm off}={n}_{\rm bulk}k^{\rm ns}_{\rm on}$ such that $K_{\rm ns}
=k_{\rm on}^{\rm ns}/k_{\rm off}^{\rm ns}$ \cite{REMact}.

To find the distribution $W_{\rm bulk}(x,t)$ for return to the DNA of a
protein released at the DNA at $x=0$ and $t=0$, we solve the cylindrical
diffusion equation for the probability density $P(x,r,t)$ of the protein's
position
\begin{equation}
\frac{\partial {P}}{\partial t}=D_{\rm
3d}\left(\frac{\partial^2 }{\partial x^2}+\frac{1}{r}\frac{\partial}{\partial
r}r\frac{\partial}{\partial r}\right){P}
\label{eq:bulkdiff}
\end{equation}
where $r$ measures the distance to the cylinder axis. The reactive boundary
condition for the probability flux out of the cylinder boundary is
\begin{equation}
-\left.2 \pi r_{\rm int} D_{\rm
3d}\frac{\partial {P}}{\partial r}\right|_{r=r_{\rm int}}=-\left.k^{\rm ns}_{\rm on}{
P}\right|_{r=r_{\rm int}}+\delta(t)\delta(x).
\label{eq:bulkbound}
\end{equation}
The initial distribution outside the cylinder is $\left.P(r,t)\right|_{t=0}=0$.
From
the solution of this problem we obtain $W_{\rm bulk}$ as $k^{\rm ns}_{\rm on}
\left.{P}\right|_{r=r_{\rm int}}$ in terms of modified Bessel functions $K_{
\nu}$ [with ${\bar q}=\sqrt{q^2+u/D_{\rm 3d}}$]
\begin{equation}
W_{\mathrm{bulk}}^{\rm cyl}(q,u)=
\left(1+\frac{2\pi D_{\rm 3d}{\bar q}r_{\rm int}K_1({\bar q} r_{\rm int})}{k^{\rm ns}_{\rm on}K_0({\bar q}
r_{\rm int})}\right)^{-1}.
\label{eq:Wbulkcyl}
\end{equation}

With these ingredients one can evaluate numerically the search rate to reach
the specific target on a straight DNA, as given by Eq.~(\ref{int}). However,
some limits allow analytic approximation for the integral in Eq.~(\ref{int}):
for $k^{\rm ns}_{\rm on}\ll D_{\rm 3d}$ one may take $1-\lambda_{\rm bulk}(q)
\approx 1$, and thus $k_{\rm 1d}^{-1}\sim\left(2\sqrt{D_{\rm 1d}k^{\rm ns}_{
\rm off}}\right)^{-1}$. One finds the association rate
\begin{equation}
\label{eq:konmax}
k_{\rm on}=2 l_{\rm sl} k^{\rm ns}_{\rm on}\;,
\end{equation}
where $l_{\rm sl}=\sqrt{D_{\rm 1d}/k^{\rm ns}_{\rm off}}$ is the so-called
sliding length of a protein during a single NSB period. Eq.~(\ref{eq:konmax})
states that each time a protein, with rate constant $k^{\rm ns}_{\rm on}$,
binds to the DNA within a distance $l_{\rm sl}$ from the target it is able
to ``slide" onto the target. Note that in this $k^{\rm ns}_{\rm on}\ll D_{\rm
3d}$ limit the search rate $k_{\rm on}$ is independent of the DNA conformation.
In the opposite limit $k^{\rm ns}_{\rm on}\gg D_{\rm 3d}$ one may approximate
$1-\lambda_{\rm bulk}(q)\approx 2\pi D_{\rm 3d}/[k_{\rm on}^{\rm ns}/\ln(l_{\rm
sl}^{\rm eff}/r_{\rm int})]$ where $l_{\rm sl}^{\rm eff}=\sqrt{k_{\rm on}^{\rm
ns}/(2\pi D_{\rm 3d})}l_{\rm sl}$. This produces the result \cite{REM2}
\begin{equation}
\label{antenna}
k_{\rm on}\sim 4 \pi D_{\rm 3d} l_{\rm sl}^{\rm eff}/[\ln(l_{\rm sl}^{\rm
eff}/r_{\rm int})]^{1/2}.
\end{equation}
In this limit the protein can rebind quickly after unbinding. This
leads to oversampling and a lowered value of $k_{\rm on}$ compared to
Eq.~(\ref{eq:konmax}). $l_{\rm sl}^{\rm eff}$ can be interpreted as
an effective sliding length including immediate rebindings \cite{berg}.
Conversely,
in the limit $k^{\rm ns}_{\rm on}\ll D_{\rm 3d}$ the rate of NSB will
limit rebinding such that the protein will relocate by a long distance during a
single 3D excursion. The part of $\lambda_{\rm bulk}$ corresponding
to such long relocations do not contribute significantly
to the integral in Eq.~(\ref{int}), which is the physical reason for the
approximation leading to Eq.~(\ref{eq:konmax}). Results (\ref{int}) and
(\ref{eq:Wbulkcyl}) are identical to the classical result for straight,
infinitely long DNA \cite{berg}. However, the approach here allows to explicitly
consider coiled DNA.

\begin{figure}
\includegraphics[width=4.8cm]{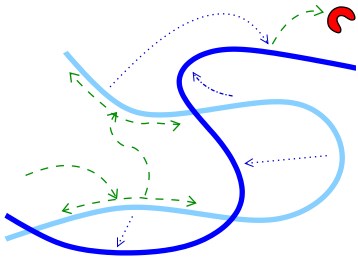}
\caption{Search rate enhancement on coiled DNA: a protein that in the case of
straight DNA would perform a short hop can now, due to DNA looping, be captured
by a foreign (remote) segment of DNA. This leads to a more efficient search
since local oversampling is reduced by these intersegmental jumps \cite{REMtransfers}. The DNA
configuration is assumed to
be dynamic such that when the protein is released from the foreign segment the
latter has moved further away from the original segment. Dashed lines: motion
of the protein; dotted lines: DNA configurational changes.}
\label{fig:coil}
\end{figure}

\emph{Coil with persistence length (Fig.~\ref{fig:coil}).}
To obtain $W_{\rm bulk}$ for a random coil DNA configuration we treat the
DNA as consisting locally of straight segments. We assume that the local
density $l_{\rm DNA}$ of DNA (length per volume) around a segment is dilute
such that the typical distance $1/\sqrt{l_{\rm DNA}}$ between segments is
much larger than $r_{\rm int}$. We then apply a superposition technique,
where the probability that the protein has reacted with a segment is
obtained by treating independently the interaction with each segment, and
then choose among these the interaction for which binding happens first. To
obtain the dynamics of the interaction of the protein with one segment of DNA
we consider a large volume $V$ around the segment of length $L$.
Placing the protein at a uniformly random position in space relative
to the DNA segment
at $t=0$ and denoting the probability that the protein has
bound to this DNA before time $t$ by $J_{\rm single}(t)$, then the probability
that it has not bound to any of $N$ pieces of DNA is (using the assumed
independent interactions) $P_{\rm surv}^{\rm foreign}(t)=\left[1-J_{\rm
single}(t)\right]^N$. Taking the limit of $V,N,L\to\infty$ for fixed
$l_{\rm DNA}=N L/V$ we get $P_{\rm surv}^{\rm foreign}(t)=\exp\left[-J_{\rm
cap}(t)\right]$, where
\begin{equation}
\frac{u^2J_{\rm cap}(u)}{k^{\rm ns}_{\rm on}l_{\rm DNA}}=
\left(1+\frac{k_{\rm on}^{\rm
ns} K_0(\sqrt{u/D_{\rm 3d}}r_{\rm int})}{2\pi\sqrt{u D_{\rm 3d}}r_{\rm int}
K_1(\sqrt{u/D_{\rm 3d}}r_{\rm int})}\right)^{-1}.
\end{equation}
$W_{\rm bulk}$ in the coiled conformation then becomes
\begin{eqnarray}
W_{\rm bulk}(q,t)&=&W_{\rm bulk}^{\rm cyl}(q,t)P_{\rm surv}^{\rm foreign}(t)
\label{eq:coilmodel}
\\
&&
-\frac{d P_{\rm surv}^{\rm foreign}(t)}{d t}\delta_{q,0}
\int_t^\infty W_{\rm bulk}^{\rm cyl}(q=0,t')d t'
\,.\nonumber
\end{eqnarray}
The first term on the right hand side stands for proteins that return to the
original segment without being captured by foreign segments. Those
are represented by the second term, where the factor $\delta_{q,0}$
is a result of the assumption that a capture by foreign DNA segments leads
to a long relocation measured along the DNA contour. These relocations only
contribute close to $q=0$ in Fourier space. Exploiting the reasoning behind
the approximation that led to Eq. (\ref{eq:konmax}) we discard them
from $W_{\rm bulk}$ when the integral in Eq. (\ref{int}) is calculated.
However, even numerically this integral is difficult to handle. A good
approximation can be found noting that by Tauberian theorems the Laplace
inversion for $J_{\rm cap}(u)$ can be divided into three regimes when
$k^{\rm ns}_{\rm on}\gg D_{\rm 3d}$:
\begin{equation}
J_{\rm cap}(t)=\left\{\begin{array}{ll} k^{\rm ns}_{\rm on}l_{\rm DNA}
t&,\; t\ll \frac{D_{\rm 3d}r_{\rm int}^2}{(k^{\rm ns}_{\rm on})^2}\\
4l_{\rm DNA}r_{\rm int}\sqrt{\pi D_{\rm 3d}t}&,\;\frac{D_{\rm 3d}r_{\rm
int}^2}{(k^{\rm ns}_{\rm on})^2}\ll t\ll \frac{r_{\rm int}^2}{D_{\rm
3d}}\\
\frac{4\pi l_{\rm DNA}D_{\rm 3d} t}{\ln (D_{\rm 3d}t/r_{\rm
int}^2)} &,\;t\gg \frac{r_{\rm int}^2}{D_{\rm 3d}}
\end{array}\right.
\end{equation}
The first regime corresponds to reaction limited binding and is very brief
for large $k^{\rm ns}_{\rm on}$. We discard this regime and ignore the
slowly varying logarithm in the last regime, to approximate $P^{\rm foreign}
_{\rm surv}$ by \cite{REM3}
\begin{equation}
\label{eq:Papprox}
P_{\rm app}(t)=\exp\left[-4 l_{\rm DNA} r_{\rm int}
\sqrt{\pi D_{\rm 3d} t}-k_{\rm cap}t\right].
\end{equation}
The value of $k_{\rm cap}$ is fixed self-consistently by the condition
that the total amount of proteins returning to
the original segment is the same as for $P_{\rm surv}^{\rm foreign}$, i.e.
\begin{eqnarray}
&&\int_0^\infty d t\;W_{\rm bulk}^{\rm cyl}(q,t)\left[P_{\rm surv}^{\rm foreign}(t)
-P_{\rm app}(t)\right]=0.
\end{eqnarray}
We solved this equation numerically. But when a value for $k_{\rm cap}$
has been obtained then the choice of $P_{\rm app}$ in
Eq. (\ref{eq:Papprox}) is very convenient.
In fact it can be written
\begin{eqnarray}
P_{\rm app}(t)&=&\int_0^\infty d s\; e^{-(s+k_{\rm cap}) t}\frac{a}{2\sqrt{
\pi s^3}}e^{-a^2/(4 s)},
\end{eqnarray}
where $a=4\sqrt{\pi}r_{\rm int}l_{\rm DNA}\sqrt{D_{\rm 3d}}$. When inserted
into Eq. (\ref{eq:coilmodel}) in Laplace space we need to evaluate numerically
only one integral to obtain $W_{\rm bulk}(q,u)$, namely
\begin{equation}
\label{res}
W_{\rm bulk}(q,u)=\int_0^\infty d s\;\frac{a\, e^{-a^2/(4 s)}}{2\sqrt{\pi s^3}}
W_{\rm bulk}^{\rm cyl}(q,u+s+k_{\rm cap}).
\end{equation}
The term with $\delta_{q,0}$ was already discarded here. Eq.~(\ref{res}) generalizes
the straight rod result
(\ref{eq:Wbulkcyl}). In the limit $D_{\rm 3d}\gg k_{\rm on}^{\rm ns}$ it is consistent with the approximation $1-\lambda_{\rm bulk}(q)\approx 1$ leading to the  expected result (\ref{eq:konmax}) for that case.

In above analysis we assumed that the density of foreign segments is uniform
in space. Although this is not true for a relaxed coil (the density decreases
on a scale of the persistence length of the DNA) we argue that this is not
important for evaluating the search rate: in fact, if the protein diffuses
far away from the original segment it will most likely perform a long
relocation measured along the DNA contour. Again it does not matter for the
search rate how we bookkeep these long relocations. To estimate the
density $l_{\rm DNA}$ for a random relaxed coil we employ the Worm Like Chain
model. A good approximation for the probability density for a point a
contour distance $s$ away from another point to loop back is \cite{ringrose99}
\begin{equation}
j_{\rm M}=\left(\frac{3}{4\pi|s|\ell_{\rm p}}\right)^{3/2}\exp\left(
-\frac{8\ell_{\rm p}^2}{|s|^2}\right).\label{eq:WLC}
\end{equation}
For a target in the middle of a chain of length $L$,
we find the DNA density $l_{\rm DNA}=\int_{-L/2}^{L/2} d s\;j_{\rm
M}$ around the target.

The theory presented here was shown in Ref.~\cite{gijs} to quantitatively
describe recent experimental data obtained from an optical tweezers setup by
which the search rate $k_{\rm on}$ at various degrees of DNA coiling could
be measured. The relative increase $R$ of $k_{\rm on}$ in the maximally
relaxed DNA configuration compared to $k_{\rm on}$ in the stretched state,
for DNA length $L=6538$ base pairs, was measured to be around 1.1 to 1.3
in buffers with salt concentrations of 5 mM ${\rm MgCl}_2$ and 0 to 25
mM for NaCl. This relative increase $R$ compared well with the theory
presented here using estimates of $l_{\rm DNA}$ based on Eq. (\ref{eq:WLC})
giving $1/\sqrt{l_{\rm DNA}}\sim 500\,{\rm bp}$, and a value $l_{\rm sl}^{\rm
eff}\sim 100\,{\rm bp}$ based on a quantitative analysis \cite{zhou05}
of experiments \cite{jeltsch98} suggesting effective sliding
lengths of this size. Numerical analysis of the theory presented here showed
that $R$ depends significantly only on $l_{\rm DNA}$ and $l_{\rm sl}^{\rm eff}$
for realistic ranges of the remaining parameters. At physiological salt
(100 mM NaCl) an increase of $R\approx1.7$ was found experimentally \cite{gijs}.
This observation agrees well with the increased DNA density expected due to
attraction between segments induced by the 5 mM divalent ${\rm Mg}^{2+}$ ions
at this NaCl concentration, see Ref.~\cite{gijs} for details. We are not aware
of a theory to calculate $l_{\rm DNA}$ under such attraction. However, the
value $1/\sqrt{l_{\rm DNA}}\sim l_{\rm sl}^{\rm eff}$ needed to obtain
$R\approx1.7$ at this salt conditions appears reasonable.

We presented a model for facilitated diffusion that is useful to calculate
the search rate of DNA binding proteins for their specific binding site on
a fluctuating DNA coil with variable DNA density. This model represents a
convenient way to rephrase the Berg-von Hippel theory with the additional
advantage that it interpolates between stretched and coiled DNA
configurations.
In the development of this
theory a number of assumptions were made, in particular that of
sufficiently fast local relaxation of the DNA configuration.
This assumption was found to be satisfied for the protein EcoRV used in
\cite{gijs}, but it will not hold universally.
It would therefore be beneficial to generalize the
theory to slowly fluctuating or static DNA configurations.

We gratefully acknowledge discussions with Tobias Ambj{\"o}rnsson, Alexander
Grosberg, Mehran Kardar, Leonid Mirny, and Roland Winkler.
This research was partially funded by the
Villum Kann Rasmussen Foundation, Deutsche Forschungsgemeinschaft, and by a
Fundamenteel Onderzoek der Materie Projectruimte Grant, a Netherlands
Organization for Scientific Research-Vernieuwings Impuls grant.


\begin{thebibliography}{99}

\bibitem{hughes} B.~D.~Hughes {\em Random Walks and Random Environments,
Vol.~1} (Oxford University Press, Oxford, 1995).

\bibitem{lomholt_pnas} M. A. Lomholt, T. Koren, R. Metzler, and J. Klafter,
Proc. Natl. Acad. Sci. USA \textbf{105}, 11055 (2008).

\bibitem{koren} T. Koren, M. A. Lomholt, A. V. Chechkin, J. Klafter, and R.
Metzler, Phys. Rev. Lett. \textbf{99}, 160602 (2007).

\bibitem{smoluchowski} M. von Smoluchowski, Physikal. Zeitschr. \textbf{17},
557 (1916).

\bibitem{benichou_pccp} O. B{\'e}nichou, C. Loverdo, M. Moreau, and R.
Voituriez, Phys. Chem. Chem. Phys. \textbf{10}, 7059 (2008).

\bibitem{bvh} P.H. von Hippel and O.G. Berg, J. Biol. Chem. \textbf{264},
675 (1989).

\bibitem{igor} I. M. Sokolov, R. Metzler, K. Pant, and M. C. Williams,
Biophys. J. \textbf{89}, 895 (2005); J. Elf, G. W. Li, and X. S. Xie,
Science \textbf{316}, 1191 (2007); Y. M. Wang, R. H. Austin, and E. C. Cox,
Phys. Rev. Lett. \textbf{97}, 048302 (2006).

\bibitem{audun} A. Bakk and R. Metzler, FEBS Lett. \textbf{563}, 66 (2004);
J. Theor. Biol. \textbf{231}, 525 (2004), and Refs. therein.

\bibitem{berg} O.~G. Berg and M. Ehrenberg, Biophys. Chem. {\bf 15}, 41
(1982).

\bibitem{REM} Even in bacteria the DNA length measures several mm while DNA's
persistence length is of the order of 50 nm.

\bibitem{hu} T. Hu, A. Y. Grosberg, and B. I. Shklovskii, Biophys. J.
\textbf{90}, 2731 (2006).

\bibitem{halford} S. E. Halford and J. F. Marko, Nucl. Acids Res. \textbf{32},
3040 (2004).

\bibitem{slutsky} M. Slutsky and L. A. Mirny, Biophys. J. \textbf{87}, 4021
(2004).

\bibitem{coppey} M. Coppey, O. B{\'e}nichou, R. Voituriez, and M. Moreau,
Biophys. J. {\bf 87}, 1640 (2004).

\bibitem{lomholt} M.~A.~Lomholt, T.~Ambj{\"o}rnsson, and R.~Metzler,
Phys.~Rev.~Lett.~\textbf{95}, 260603 (2005).

\bibitem{gijs} B. van den Broek, M. A. Lomholt, S.-M. J. Kalisch, R. Metzler,
and G. J. L. Wuite, Proc. Natl. Acad. Sci. USA {\bf 105}, 15738 (2008).

\bibitem{REMinf} In practice the DNA should be longer than the effective sliding length
$l_{\rm sl}^{\rm eff}$ introduced
before Eq. (\ref{antenna}).

\bibitem{REMact}We deviate here from conventions in \cite{gijs} by only letting $n_{\rm bulk}$ count `active enzymes'.

\bibitem{REM2} Eq.~(\ref{antenna}) generalizes Smoluchowski's result $k_{
\mathrm{on}}=4\pi D_{\mathrm{3d}}b$ for pure 3D search for a target of size
$b$. The effective target size in Eq.~(\ref{antenna}) is often called
antenna length.

\bibitem{REMtransfers}Note the distinction to intersegmental transfers where the protein intermittently is bound to two DNA segments \cite{bvh}.

\bibitem{REM3} $P_{\mathrm{surv}}^{\mathrm{foreign}}$ and $P_{\mathrm{app}}$
agree very well numerically.

\bibitem{ringrose99}L. Ringrose {\emph{et al.}}, EMBO J. {\bf 18}, 6630 (1999).

\bibitem{zhou05}H.-X. Zhou, Biophys. J. {\bf 88}, 1608 (2005).

\bibitem{jeltsch98}A. Jeltsch and A. Pingoud, Biochemistry {\bf 37}, 2160 (1998);
N.~P. Stanford {\emph{et al.}},
EMBO J. {\bf 19}, 6546 (2000).

\end{thebibliography}
\end{document}